RWTH Aachen

Department of Computer Science
*Technical Report*

# An Operational Semantics for Activity Diagrams using SMV

Shahar Maoz and Jan Oliver Ringert and Bernhard Rumpe

RWTH Aachen
Software Engineering

# An Operational Semantics for Activity Diagrams using SMV

Shahar Maoz[1]
Jan Oliver Ringert[2]
Bernhard Rumpe

Aachen, July 12, 2011

[1]S. Maoz acknowledges support from a postdoctoral Minerva Fellowship, funded by the German Federal Ministry for Education and Research.
[2]J.O. Ringert is supported by the DFG GK/1298 AlgoSyn.

**Abstract**

This document defines an operational semantics for activity diagrams (ADs) using a translation to SMV. The translation is inspired by the work of Eshuis [Esh06] and extends it with support for data. Each execution step of the SMV module obtained from an AD represents an executed action of this AD with interleaved execution of concurrent branches.

An implementation of the given translation was used in the context of semantic differencing for ADs [MRR11]. We define the translation and give two examples, showing ADs and their complete representation in SMV.

# Chapter 1

# Activity Diagrams

Activity diagrams (ADs) have recently become widely used in the modeling of work flows, business processes, and web-services, where they serve various purposes, from documentation, requirement definitions, and test case specifications, to simulation and code generation.

## 1.1 AD Language Syntax

An Activity Diagram is a structure
$AD = \langle A, V^{inp}, V^{loc}, AN, PN, T\rangle$ where:

- $A$ is a set of action names.
- $V^{inp}$ is a (possibly empty) set of immutable input variables over finite domains.
- $V^{loc}$ is a (possibly empty) set of local variables over finite domains.
- $AN$ is a set of action nodes $an_1, \ldots, an_k$. Each action node $an$ is labeled with an action name $acname(an) = ac \in A$, and a (possibly empty) set of assignment expressions to the variables in $V^{loc}$.
- $PN$ is a set of pseudo nodes, consisting of initial nodes $PN^{init}$, final nodes $PN^{fin}$, decision nodes $PM^{dec}$, merge nodes $PN^{mer}$, fork nodes $PN^{fork}$, and join nodes $PN^{join}$.
- $T$ is a set of transitions of the form $t = \langle n_{src}, n_{tgt}, guard\rangle$ where $n_{src}, n_{tgt} \in (AN \cup PN)$ and $guard$ is a Boolean expression over the variables in $V^{inp} \cup V^{loc}$. Unless $n_{src}$ is a decision node, $guard = \texttt{true}$.

We do not formally capture here obvious well-formedness rules and context conditions such as: initial nodes have no incoming transitions, final nodes have no outgoing transitions, etc.

A minor technical limitation of our current translation rules is that two pseudo nodes should not be connected directly. Some rules for translating pseudo nodes rely on their successors or predecessors to determine the previous or next action. This limitation can be easily removed by 'skipping' pseudo nodes when looking for the next action in the abstract syntax.

## 1.2 Operational Semantics

We give operational semantics to activity diagrams using a translation to SMV, the language of the SMV model checker [SMV]. The translation is inspired by the translation presented in [Esh06]. It extends this previous translation with support for data.

The SMV language allows the description of finite state machines (FSMs). FSMs consist of a set of variables and predicates on these variables. Predicates use the logical operators `&` (and), `|` (or), and `!` (not). Constant `1` denotes true whereas `0` denotes false. Variables are declared using the `VAR` keyword, followed by a list of typed variable declarations. Variables can be of type Boolean or can be enumerative. A state is an assignment of values to a set of variables. Predicates are of two types: predicates defining the initial state are preceded by the `INIT` keyword, and predicates defining the transition relation, relating the current values of some variables with their possible next values, are preceded by the `TRANS` keyword.

The SMV language was presented in [McM93]. Complete syntax and semantics definitions for SMV can be found in [CCJ+05].

We present our complete translation of ADs given in abstract syntax to the SMV language in Figures 1.1, 1.2 and 1.3. An AD is translated into one SMV module, i.e., FSM. Each step of this FSM represents the execution of a single action of the AD. Termination of the AD is represented by an infinite execution sequence of the pseudo action `nop`.

Rule 1 from Fig. 1.1 gives the state space of the FSM. It consists of control flow variables for each node that show if this node is active or not. Fork nodes are translated as separate variables for each outgoing transition and join nodes as separate variables for incoming transitions. In all cases these variables are used to decide which transition steps can be executed (see rule 3). Variable `acnode` denotes the action node in each step and variable `ac` holds the name of the executed action. Local and initial variables are translated directly to variables in SMV.

1 VAR

$\forall an \in AN \cup PN^{init,fin}$ :

    in_$an.nId$: boolean;

$\forall fn \in PN^{fork} \forall t \in fn.out$ :

    in_F$t.tgt.nId$: boolean;

$\forall jn \in PN^{join} \forall t \in jn.in$ :

    in_J$t.src.nId$: boolean;

acnode : $\{\bigcup_{an \in AN} an.nId\}$;

ac : $\{\bigcup_{acname \in A} acname\}$;

$\forall var \in (V^{inp} \cup V^{loc})$ :

    $var.name$ : $var.typeDecl$;

2 INIT

in_$ad.initialNode.nId$ = 1 &

$\forall an \in AN$ :

    in_$an.nId$ = 0 &

$\forall fn \in PN^{fork} \forall t \in fn.out$ :

    $\bigwedge$ in_F$t.tgt.nId$ = 0 &

$\forall jn \in PN^{join} \forall t \in jn.in$ :

    $\bigwedge$ in_J$t.src.nId$ = 0 &

$\forall var \in V^{loc}$ :

    $var.name$ = $var.init$ &

acnode = $ad.initialNode.nId$ &

ac = $ad.initialNode.acName$;

Figure 1.1: AD FSM variables and initial states

Rule 2 specifies all initial states. The control flow variable of the initial node is marked true, and all others false. Local variables are initialized to their pre-defined values. The value of acnode and ac is determined by the initial node. Input variables are not assigned a value (they get their value from the environment).

Rule 3 from Fig. 1.2 defines taking of transitions: a transition's source node has to be active and its (optional) guard needs to evaluate to true. When a transition is taken, $t$_`taken` disables predecessor nodes, enables successor control flow variables in the next state, and the variable `acnode` is updated to the next action node. Special definitions, $3^f$, $3^j$,$3^m$ and $3^d$, handle pseudo nodes where diagram edges do not relate one-to-one to state transitions.

Rule $3^f$ states that every transition $t$ leaving a fork node can be taken if it's control flow variable `in_F`$t.tgt.nId$ is true. These variables are set in rule 4. To take a transition preceding a join node, all control flow variables `in_J`$t'.src.nId$ have to be true, indicating that all previous concurrent branches have reached the join node (see rule $3^j$).

Transitions to merge nodes are routed to the target node of the one outgoing edge of the merge (rule $3^m$). A transition leaving a decision node can be taken if its guard evaluates to true and the control flow has reached the node preceding it (rule $3^d$).

3 DEFINE

$\forall t \in T \wedge t.src, t.tgt \in AN \cup PN^{initial,final}$ :
$t$_taken := in_$t.src.nId$ &
!next(in_$t.src.nId$) &
next(in_$t.tgt.nId$) &
next(acnode = $t.tgt.nId$);

$3^f$ $\forall t \in T \wedge t.src \in PN^{fork}$ :
$t$_taken := in_F$t.tgt.nId$ &
!next(in_$t.src.in.src.nId$) &
!next(in_F$t.tgt.nId$) &
next(in_$t.tgt.nId$) &
next(acnode = $t.tgt.nId$);

$3^j$ $\forall t \in T \wedge t.src \in PN^{join}$ :
$t$_taken := $\bigwedge_{t' \in t.src.in}$ in_J$t'.src.nId$ &
$\bigwedge_{t' \in t.src.in}$ !next(in_J$t'.src.nId$) &
next(in_$t.tgt.nId$) &
next(acnode = $t.tgt.nId$);

$3^m$ $\forall t \in T \wedge t.tgt \in PN^{mer}$ :
$t$_taken := in_$t.src.nId$ &
!next(in_$t.src.nId$) &
next(in_$t.tgt.out.tgt.nId$) &
next(acnode = $t.tgt.out.tgt.nId$);

$3^d$ $\forall t \in T \wedge t.src \in PN^{dec}$ :
$t$_taken := in_$t.src.in.src.nId$ &
$t.guard$ &
!next(in_$t.src.in.src.nId$) &
next(in_$t.tgt.nId$) &
next(acnode = $t.tgt.nId$);

Figure 1.2: AD FSM transition definitions

Rule 4 activates the control flow variables for forking if the next step arrives in an action node previous to the fork and a corresponding join variable when executing an action before the join node.

Rule 5 defines that by every state transition the control flow variables `in_`$an.nId$ are not changed, unless incoming or outgoing edges are taken. Fork nodes can change these variables with every outgoing transition and join nodes with their one outgoing transition. The special variables `in_F`$t.tgt.nId$ for fork and `in_J`$t.src.nId$ join nodes are changed if incoming or outgoing edges from or to these pseudo nodes are taken.

Rule 6 ensures that – unless the diagram traversal has reached a final node – in every step of the SMV FSM one edge of the diagram has to be taken. With $t$`_taken`'s unique assignments to `acnode`, this results in exactly one following action node.

Rule 7 states that during every transition the value of input variables stays constant.

Rule 8 is used to specify that local variables can only change to the next step if the next node contains an assignment to a variable with this name. If so the variable will have the assigned value in the next step.

Rule 9 assigns the executed action's name for the given action node `acnode` to variable `ac`.

4 TRANS

$\forall fn \in PN^{fork}, n = fn.in.src :$
$\quad (\ \texttt{next(acnode)} = n.nodeId \texttt{ -> }$
$\qquad \bigwedge_{t \in fn.out} \texttt{next(in_F}t.tgt.nId\texttt{)}\ )\ \texttt{\&}$
$\forall jn \in PN^{join} \forall t \in jn.in :$
$\quad \bigwedge(\ \texttt{next(acnode)} = t.src.nodeId \texttt{ -> }$
$\qquad \texttt{next(in_J}t.src.nId\texttt{)}\ );$

5 $\forall n \in AN :$
$\quad (\ \texttt{in_}n.nId = \texttt{next(in_}n.nId\texttt{)}\ |$
$\quad \bigvee_{t \in n.in \cup n.out, t\texttt{_taken defined}} t\texttt{_taken}\ |$
$\quad \bigvee_{t \in n.out.tgt.out, t.src \in PN^{fork}} t\texttt{_taken}\ |$
$\quad \bigvee_{t = n.out.tgt.out, t.src \in PN^{join}} t\texttt{_taken}\ )\ \texttt{\&}$
$\forall fn \in PN^{fork} \forall t \in fn.out :$
$\quad (\ \texttt{in_F}t.tgt.nId = \texttt{next(in_F}t.tgt.nId\texttt{)}\ |$
$\quad \bigvee_{t' \in fn.in.src.in \cup t} t'\texttt{_taken}\ )\ \texttt{\&}$
$\forall jn \in PN^{join} \forall t \in jn.in :$
$\quad (\ \texttt{in_J}t.src.nId = \texttt{next(in_J}t.src.nId\texttt{)}\ |$
$\quad \bigvee_{t' \in t.src.in \cup jn.out} t'\texttt{_taken}\ );$

6 $\forall n \in PN^{final} :$
$\quad \bigwedge\ \texttt{in_}n.nId \texttt{ -> next(acnode = nop)}\ \texttt{\&}$
$\quad (\ \bigvee\ \texttt{in_}n.nId\ |$
$\qquad \bigvee_{t \in T, t\texttt{_taken defined}} t\texttt{_taken}\ );$

7 $\forall var \in V^{inp} : \bigwedge var.vName = \texttt{next(}var.vName\texttt{)};$

8 $\forall v \in V^{loc} :$
$\quad \bigwedge(\ v.vName = \texttt{next(}v.vName\texttt{)}\ |$
$\qquad \bigvee_{n \in asgnVar(v)} (\texttt{next(acnode)} = n.nId\ \texttt{\&}$
$\qquad\quad \texttt{next(}v.vName\texttt{)} = n.asgmt_v.val)\ );$

9 $\bigwedge_{an \in AN} (\texttt{next(acnode)} = an.nId \texttt{ -> }$
$\qquad \texttt{next(ac)} = an.acName);$

Figure 1.3: AD FSM transition rules

# Chapter 2

# Examples of AD to SMV Transformation

We present two complete examples of the translation of an activity diagram to SMV code. The first activity from Fig. 2.1 contains internal control and external input variables. Its action nodes contain assignments to local variables. A decision node evaluates expressions over internal and external variabels. The second example in Fig. 2.2 shows an activity where actions are executed in parallel (interleaved) with nondeterministic choice of their execution order.

## 2.1 Example I

The AD `controlledLoop` from Fig. 2.1 contains four action nodes, an input variable `project` with values `short` and `long`, and an internal local variable `iterations` with domain {`0, 1, 2, 3, 4`}. Variable `iterations` is initialized to `0` in the first action node. Its value is incremented each time action `work` is executed. The loop containing the two actions `define work` and `work` can only be left if the input variable `project` was initially set to `short` by the environemt or the local variable `iterations` has been increased to 3 after executing the loop `define work` and `work` three times.

The fully automated translation to SMV code following the scheme presented above in Fig. 1.1 and Fig. 1.3 consists of about 160 lines of SMV code presented in listings 2.1, 2.2, 2.3, 2.4 and 2.5. Nodes cannot be identified by their action names only since these might be used more than once in an activity. The nodes in this example are `n1` to `n4` representing the action nodes from Fig. 2.1 and the special nodes `n0_initial` and `n5_final` for the initial and final node (see l.11 of listing 2.1). The correspondance of nodes

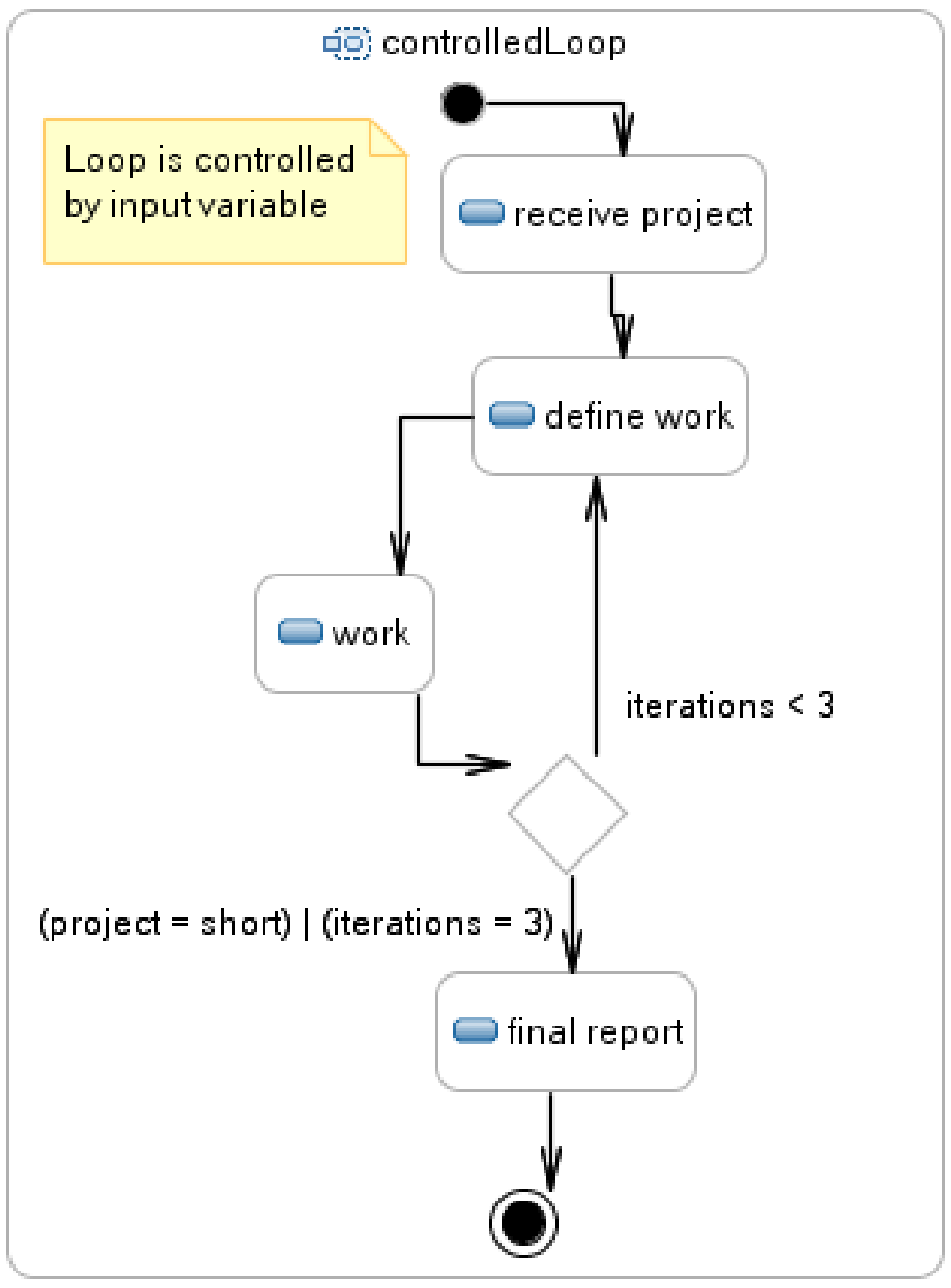

Figure 2.1: Activity diagram `controlledLoop`

and actions is established in listing 2.5 as defined in rule 9 of Fig. 1.3.

It is also easy to spot the application of rule $3^d$ from Fig. 1.1 which is applied to all edges that are outgoing from the decision node in the AD. The edge with guard `( iterations < 3 )` leading to action `define work` produces lines 28-38 in listing 2.2. Please note that in the generated code the first condition of whether edges can be taken or not is always factored out into an additional definition ending with `_enabled`. The second edge leaving the decision node with guard `( ( project = short ) | ( iterations = 3) )` produces accordingly lines 1-12 in listing 2.3.

```
VAR
  -- nodes and pseudo-nodes of ad
  in_n0_initial : boolean;
  in_n1 : boolean;
  in_n2 : boolean;
  in_n3 : boolean;
  in_n4 : boolean;
  in_n5_final : boolean;

  -- visitable nodes
  acnode : {n0_initial, n1, n2, n3, n4, n5_final, nop};

  -- the visible action of a step
  ac : {define_work, final_report, receive_project,
                                                work, nop};

  -- input variables
  project : {long, short};

  -- control variables
  iterations : {0,1,2,3,4};

INIT
  -- init all nodes
  in_n0_initial = 1 &
  in_n1 = 0 &
  in_n2 = 0 &
  in_n3 = 0 &
  in_n4 = 0 &
  in_n5_final = 0 &
  -- init control variables as assigned in first node
  iterations = ( 0) &
  -- set initial action node and visible action
  acnode = n0_initial &
  ac = nop;
```

Listing 2.1: Variables and their initial values of automaton of AD `controlledLoop`

```
-- shortcut to what happens when an edge is taken
DEFINE
  en0_initialn1_enabled := in_n0_initial ;
  en0_initialn1_taken := en0_initialn1_enabled &
    -- not in previous nodes anymore
    !next(in_n0_initial) &
    -- arrive in target node
    next(in_n1) &
    -- possibly taking hidden edges
    -- doing assignments
    next (iterations) =  0 &
    -- set next node
    next(acnode = n1);

DEFINE
  en2n3_enabled := in_n2 ;
  en2n3_taken := en2n3_enabled &
    -- not in previous nodes anymore
    !next(in_n2) &
    -- arrive in target node
    next(in_n3) &
    -- possibly taking hidden edges
    -- doing assignments
    next (iterations) =  iterations +1 &
    -- set next node
    next(acnode = n3);

DEFINE
  en3n2_enabled := in_n3  & (iterations < 3);
  en3n2_taken := en3n2_enabled &
    -- not in previous nodes anymore
    !next(in_n3) &
    -- arrive in target node
    next(in_n2) &
    -- possibly taking hidden edges
    -- doing assignments
    -- set next node
    next(acnode = n2);
```

Listing 2.2: Shortcuts to define what happens when edges are taken (part 1)

```
DEFINE
  en3n4_enabled := in_n3  &
    ((project = short) | (iterations = 3));
  en3n4_taken := en3n4_enabled &
    -- not in previous nodes anymore
    !next(in_n3) &
    -- arrive in target node
    next(in_n4) &
    -- possibly taking hidden edges
    -- doing assignments
    -- set next node
    next(acnode = n4);

DEFINE
  en4n5_final_enabled := in_n4 ;
  en4n5_final_taken := en4n5_final_enabled &
    -- not in previous nodes anymore
    !next(in_n4) &
    -- arrive in target node
    next(in_n5_final) &
    -- possibly taking hidden edges
    -- doing assignments
    -- set next node
    next(acnode = n5_final);

DEFINE
  en1n2_enabled := in_n1 ;
  en1n2_taken := en1n2_enabled &
    -- not in previous nodes anymore
    !next(in_n1) &
    -- arrive in target node
    next(in_n2) &
    -- possibly taking hidden edges
    -- doing assignments
    -- set next node
    next(acnode = n2);
```

Listing 2.3: Shortcuts to define what happens when edges are taken (part 2)

```
TRANS
  (  (in_n0_initial = next(in_n0_initial)  ) |
      en0_initialn1_taken ) &
  (  (in_n1 = next(in_n1)  ) |
      en0_initialn1_taken |
      en1n2_taken ) &
  (  (in_n2 = next(in_n2)  ) |
      en3n2_taken |
      en1n2_taken |
      en2n3_taken ) &
  (  (in_n3 = next(in_n3)  ) |
      en2n3_taken |
      en3n2_taken |
      en3n4_taken ) &
  (  (in_n4 = next(in_n4)  ) |
      en3n4_taken |
      en4n5_final_taken ) &
  (  (in_n5_final = next(in_n5_final)  ) |
      en4n5_final_taken );

TRANS
  (   (next(acnode=nop) <-> in_n5_final  ) ) &
      ( in_n5_final  |   (
        (en0_initialn1_taken) |
        (en2n3_taken) |
        (en3n2_taken) |
        (en3n4_taken) |
        (en4n5_final_taken) |
        (en1n2_taken) ))  ;

TRANS
-- input variables do not change
  project = next(project) ;

TRANS
  -- local variables change only on assignments
  (  iterations = next(iterations)
   | next(acnode) = n1  | next(acnode) = n3 );
```

Listing 2.4: Transitions of automaton generated for AD `controlledLoop` (part 1)

```
1 TRANS
2   (next(acnode) = n0_initial -> next(ac) = nop )&
3   (next(acnode) = n1 -> next(ac) = receive_project )&
4   (next(acnode) = n2 -> next(ac) = define_work )&
5   (next(acnode) = n3 -> next(ac) = work )&
6   (next(acnode) = n4 -> next(ac) = final_report )&
7   (next(acnode) = n5_final -> next(ac) = nop )&
8   (next(acnode) = nop -> next(ac) = nop );
```

Listing 2.5: Transitions of automaton generated for AD `controlledLoop` (part 2)

## 2.2 Example II

The AD `hireEmployeeSimplified` from Fig. 2.2 contains four action nodes. The modeled activity shows the simplified process of hiring an employee. It starts with the action `register` of registering the new employee in the office. The control flow then forks and the actions `assign to project` and `add to website` are executed in parallel. After execution both the control flow merges and the action `authorize payment` is executed before the final node of the AD is reached.

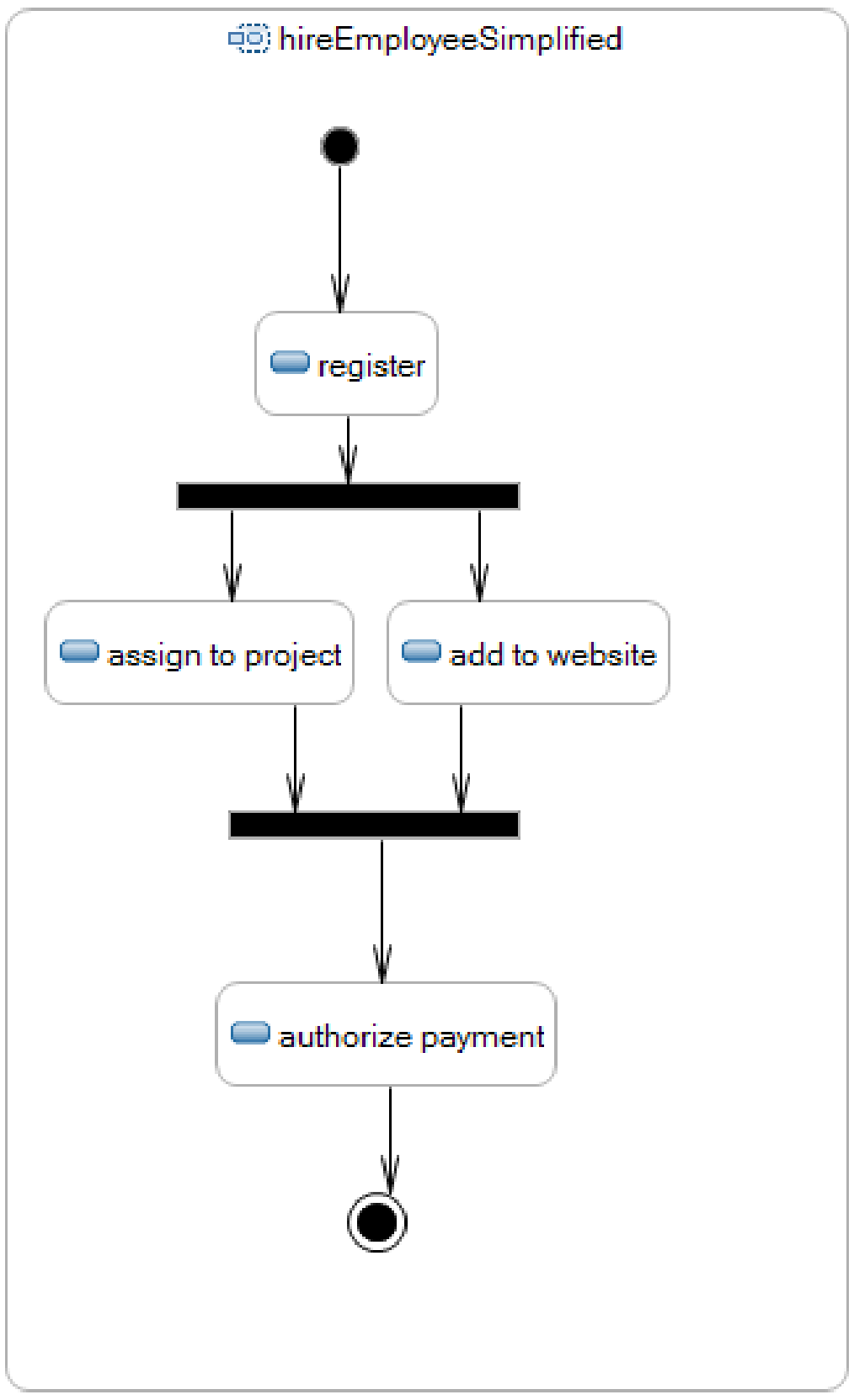

Figure 2.2: Activity diagram `hireEmployeeSimplified`

The fully automated translation to SMV code following the scheme presented above in Fig. 1.1 and Fig. 1.3 consists of about 160 lines of SMV code presented in listings 2.6, 2.7, 2.8, 2.9 and 2.10.

Listing 2.6 shows that the variables `acnode` and `ac` are generated similar as in the previous example. Additionally variables for each transition leaving a fork node are generated (see lines 9-10). These variables are set to `true` when the last action before the fork node is executed (see lines 10-11, listing 2.7).

Similar variables in listing 2.6 are generated for all predecessors of join nodes (see lines 11-12). These are each enabled after their corresponding action before the join node is executed (see lines 10 and 24, listing 2.8). The action after the join can only be executed if all of these join variables are set to `true`, i.e., all concurrent control flow paths have reached the join node (see line 17, listing 2.7).

```
VAR
  -- nodes and pseudo-nodes of ad
  in_n0_initial : boolean;
  in_n1_final : boolean;
  in_n2 : boolean;
  in_n3 : boolean;
  in_n4 : boolean;
  in_n5 : boolean;
  in_Fn4 : boolean;
  in_Fn3 : boolean;
  in_Jn3 : boolean;
  in_Jn4 : boolean;

  -- visitable nodes
  acnode : {n0_initial, n1_final, n2, n3, n4, n5, nop};

  -- the visible action of a step
  ac : {add_to_website , assign_to_project ,
                authorize_payment , nop , register };

  -- input variables

  -- control variables

INIT
  -- init all nodes
  in_n0_initial = 1 &
  in_n1_final = 0 &
  in_n2 = 0 &
  in_n3 = 0 &
  in_n4 = 0 &
  in_n5 = 0 &
  in_Fn4 = 0 &
  in_Fn3 = 0 &
  in_Jn3 = 0 &
  in_Jn4 = 0 &
  -- init control variables as assigned in first node
  -- set initial visible action node and visible action
  acnode = n0_initial &
  ac = nop;
```

Listing 2.6: Variables and their initial values of automaton of AD `hireEmployeeSimplified`

```
-- shortcut to what happens when an edge is taken
DEFINE
  en0_initialn2_enabled := in_n0_initial ;
  en0_initialn2_taken := en0_initialn2_enabled &
    -- not in previous nodes anymore
    !next(in_n0_initial) &
    -- arrive in target node
    next(in_n2) &
    -- possibly taking hidden edges
    next(in_Fn3) &
    next(in_Fn4) &
    -- doing assignments
    -- set next node
    next(acnode = n2);

DEFINE
  eJn3Jn4n5_enabled := in_Jn3 & in_Jn4 ;
  eJn3Jn4n5_taken := eJn3Jn4n5_enabled &
    -- not in previous nodes anymore
    !next(in_Jn3) &
    !next(in_n3) &
    !next(in_Jn4) &
    !next(in_n4) &
    -- arrive in target node
    next(in_n5) &
    -- possibly taking hidden edges
    -- doing assignments
    -- set next node
    next(acnode = n5);

DEFINE
  en5n1_final_enabled := in_n5 ;
  en5n1_final_taken := en5n1_final_enabled &
    -- not in previous nodes anymore
    !next(in_n5) &
    -- arrive in target node
    next(in_n1_final) &
    -- possibly taking hidden edges
    -- doing assignments
    -- set next node
    next(acnode = n1_final);
```
Listing 2.7: Shortcuts to define what happens when edges are taken (part 1)

```
DEFINE
  eFn3n3_enabled := in_Fn3 ;
  eFn3n3_taken := eFn3n3_enabled &
    -- not in previous nodes anymore
    !next(in_Fn3) &
    !next(in_n2) &
    -- arrive in target node
    next(in_n3) &
    -- possibly taking hidden edges
    next(in_Jn3) &
    -- doing assignments
    -- set next node
    next(acnode = n3);

DEFINE
  eFn4n4_enabled := in_Fn4 ;
  eFn4n4_taken := eFn4n4_enabled &
    -- not in previous nodes anymore
    !next(in_Fn4) &
    !next(in_n2) &
    -- arrive in target node
    next(in_n4) &
    -- possibly taking hidden edges
    next(in_Jn4) &
    -- doing assignments
    -- set next node
    next(acnode = n4);
```
Listing 2.8: Shortcuts to define what happens when edges are taken (part 2)

```
TRANS
  (  (in_n0_initial = next(in_n0_initial)  ) |
      en0_initialn2_taken ) &
  (  (in_n1_final = next(in_n1_final)  ) |
      en5n1_final_taken ) &
  (  (in_n2 = next(in_n2)  ) |
      en0_initialn2_taken |
      eFn3n3_taken |
      eFn4n4_taken ) &
  (  (in_n3 = next(in_n3)  ) |
      eFn3n3_taken |
      eJn3Jn4n5_taken ) &
  (  (in_n4 = next(in_n4)  ) |
      eFn4n4_taken |
      eJn3Jn4n5_taken ) &
  (  (in_n5 = next(in_n5)  ) |
      eJn3Jn4n5_taken |
      en5n1_final_taken ) &
  (  (in_Fn4 = next(in_Fn4)  ) |
      en0_initialn2_taken |
      eFn4n4_taken ) &
  (  (in_Fn3 = next(in_Fn3)  ) |
      en0_initialn2_taken |
      eFn3n3_taken 0 ) &
  (  (in_Jn3 = next(in_Jn3)  ) |
      eFn3n3_taken |
      eJn3Jn4n5_taken ) &
  (  (in_Jn4 = next(in_Jn4)  ) |
      eFn4n4_taken |
      eJn3Jn4n5_taken );

TRANS
  (   (next(acnode=nop) <-> in_n1_final  ) ) &
      ( in_n1_final  |   (
        (en0_initialn2_taken) |
        (eJn3Jn4n5_taken) |
        (en5n1_final_taken) |
        (eFn3n3_taken) |
        (eFn4n4_taken) ))  ;
```

Listing 2.9: Transitions of automaton generated for AD `hireEmployeeSimplified` (part 1)

```
1  TRANS
2    (next(acnode) = n0_initial -> next(ac) = nop)&
3    (next(acnode) = n1_final -> next(ac) = nop)&
4    (next(acnode) = n2 -> next(ac) = register)&
5    (next(acnode) = n3 -> next(ac) = assign_to_project)&
6    (next(acnode) = n4 -> next(ac) = add_to_website)&
7    (next(acnode) = n5 -> next(ac) = authorize_payment)&
8    (next(acnode) = nop -> next(ac) = nop);
```

Listing 2.10: Transitions of automaton generated for AD `hireEmployeeSimplified` (part 2)

# Bibliography

[CCJ+05] Roberto Cavada, Alessandro Cimatti, Charles Arthur Jochim, Gavin Keighren, Emanuele Olivetti, Marco Pistore, Marco Roveri, and Andrei Tchaltsev. NuSMV User Manual, 2005.

[Esh06] Rik Eshuis. Symbolic model checking of UML activity diagrams. *ACM Trans. Softw. Eng. Methodol.*, 15(1):1–38, 2006.

[McM93] K. McMillan. *Symbolic Model Checking*. Kluwer Academic Publishers, 1993.

[MRR11] Shahar Maoz, Jan Oliver Ringert, and Bernhard Rumpe. ADDiff: Semantic differencing for activity diagrams. In *ESEC/FSE'11*. ACM, 2011.

[SMV] SMV model checker. http://www.cs.cmu.edu/~modelcheck/smv.html.

## Aachener Informatik-Berichte

**This is the list of all technical reports since 1987. To obtain copies of reports please consult http://aib.informatik.rwth-aachen.de/**
**or send your request to: Informatik-Bibliothek, RWTH Aachen, Ahornstr. 55, 52056 Aachen, Email:** `biblio@informatik.rwth-aachen.de`

1987-01 * Fachgruppe Informatik: Jahresbericht 1986
1987-02 * David de Frutos Escrig, Klaus Indermark: Equivalence Relations of Non-Deterministic Ianov-Schemes
1987-03 * Manfred Nagl: A Software Development Environment based on Graph Technology
1987-04 * Claus Lewerentz, Manfred Nagl, Bernhard Westfechtel: On Integration Mechanisms within a Graph-Based Software Development Environment
1987-05 * Reinhard Rinn: Über Eingabeanomalien bei verschiedenen Inferenzmodellen
1987-06 * Werner Damm, Gert Döhmen: Specifying Distributed Computer Architectures in AADL*
1987-07 * Gregor Engels, Claus Lewerentz, Wilhelm Schäfer: Graph Grammar Engineering: A Software Specification Method
1987-08 * Manfred Nagl: Set Theoretic Approaches to Graph Grammars
1987-09 * Claus Lewerentz, Andreas Schürr: Experiences with a Database System for Software Documents
1987-10 * Herbert Klaeren, Klaus Indermark: A New Implementation Technique for Recursive Function Definitions
1987-11 * Rita Loogen: Design of a Parallel Programmable Graph Reduction Machine with Distributed Memory
1987-12 J. Börstler, U. Möncke, R. Wilhelm: Table compression for tree automata
1988-01 * Gabriele Esser, Johannes Rückert, Frank Wagner Gesellschaftliche Aspekte der Informatik
1988-02 * Peter Martini, Otto Spaniol: Token-Passing in High-Speed Backbone Networks for Campus-Wide Environments
1988-03 * Thomas Welzel: Simulation of a Multiple Token Ring Backbone
1988-04 * Peter Martini: Performance Comparison for HSLAN Media Access Protocols
1988-05 * Peter Martini: Performance Analysis of Multiple Token Rings
1988-06 * Andreas Mann, Johannes Rückert, Otto Spaniol: Datenfunknetze
1988-07 * Andreas Mann, Johannes Rückert: Packet Radio Networks for Data Exchange
1988-08 * Andreas Mann, Johannes Rückert: Concurrent Slot Assignment Protocol for Packet Radio Networks
1988-09 * W. Kremer, F. Reichert, J. Rückert, A. Mann: Entwurf einer Netzwerktopologie für ein Mobilfunknetz zur Unterstützung des öffentlichen Straßenverkehrs
1988-10 * Kai Jakobs: Towards User-Friendly Networking
1988-11 * Kai Jakobs: The Directory - Evolution of a Standard
1988-12 * Kai Jakobs: Directory Services in Distributed Systems - A Survey
1988-13 * Martine Schümmer: RS-511, a Protocol for the Plant Floor

1988-14 * U. Quernheim: Satellite Communication Protocols - A Performance Comparison Considering On-Board Processing

1988-15 * Peter Martini, Otto Spaniol, Thomas Welzel: File Transfer in High Speed Token Ring Networks: Performance Evaluation by Approximate Analysis and Simulation

1988-16 * Fachgruppe Informatik: Jahresbericht 1987

1988-17 * Wolfgang Thomas: Automata on Infinite Objects

1988-18 * Michael Sonnenschein: On Petri Nets and Data Flow Graphs

1988-19 * Heiko Vogler: Functional Distribution of the Contextual Analysis in Block-Structured Programming Languages: A Case Study of Tree Transducers

1988-20 * Thomas Welzel: Einsatz des Simulationswerkzeuges QNAP2 zur Leistungsbewertung von Kommunikationsprotokollen

1988-21 * Th. Janning, C. Lewerentz: Integrated Project Team Management in a Software Development Environment

1988-22 * Joost Engelfriet, Heiko Vogler: Modular Tree Transducers

1988-23 * Wolfgang Thomas: Automata and Quantifier Hierarchies

1988-24 * Uschi Heuter: Generalized Definite Tree Languages

1989-01 * Fachgruppe Informatik: Jahresbericht 1988

1989-02 * G. Esser, J. Rückert, F. Wagner (Hrsg.): Gesellschaftliche Aspekte der Informatik

1989-03 * Heiko Vogler: Bottom-Up Computation of Primitive Recursive Tree Functions

1989-04 * Andy Schürr: Introduction to PROGRESS, an Attribute Graph Grammar Based Specification Language

1989-05 J. Börstler: Reuse and Software Development - Problems, Solutions, and Bibliography (in German)

1989-06 * Kai Jakobs: OSI - An Appropriate Basis for Group Communication?

1989-07 * Kai Jakobs: ISO's Directory Proposal - Evolution, Current Status and Future Problems

1989-08 * Bernhard Westfechtel: Extension of a Graph Storage for Software Documents with Primitives for Undo/Redo and Revision Control

1989-09 * Peter Martini: High Speed Local Area Networks - A Tutorial

1989-10 * P. Davids, Th. Welzel: Performance Analysis of DQDB Based on Simulation

1989-11 * Manfred Nagl (Ed.): Abstracts of Talks presented at the WG '89 15th International Workshop on Graphtheoretic Concepts in Computer Science

1989-12 * Peter Martini: The DQDB Protocol - Is it Playing the Game?

1989-13 * Martine Schümmer: CNC/DNC Communication with MAP

1989-14 * Martine Schümmer: Local Area Networks for Manufactoring Environments with hard Real-Time Requirements

1989-15 * M. Schümmer, Th. Welzel, P. Martini: Integration of Field Bus and MAP Networks - Hierarchical Communication Systems in Production Environments

1989-16 * G. Vossen, K.-U. Witt: SUXESS: Towards a Sound Unification of Extensions of the Relational Data Model

1989-17 * J. Derissen, P. Hruschka, M.v.d. Beeck, Th. Janning, M. Nagl: Integrating Structured Analysis and Information Modelling
1989-18 A. Maassen: Programming with Higher Order Functions
1989-19 * Mario Rodriguez-Artalejo, Heiko Vogler: A Narrowing Machine for Syntax Directed BABEL
1989-20 H. Kuchen, R. Loogen, J.J. Moreno Navarro, M. Rodriguez Artalejo: Graph-based Implementation of a Functional Logic Language
1990-01 * Fachgruppe Informatik: Jahresbericht 1989
1990-02 * Vera Jansen, Andreas Potthoff, Wolfgang Thomas, Udo Wermuth: A Short Guide to the AMORE System (Computing Automata, MOnoids and Regular Expressions)
1990-03 * Jerzy Skurczynski: On Three Hierarchies of Weak SkS Formulas
1990-04 R. Loogen: Stack-based Implementation of Narrowing
1990-05 H. Kuchen, A. Wagener: Comparison of Dynamic Load Balancing Strategies
1990-06 * Kai Jakobs, Frank Reichert: Directory Services for Mobile Communication
1990-07 * Kai Jakobs: What's Beyond the Interface - OSI Networks to Support Cooperative Work
1990-08 * Kai Jakobs: Directory Names and Schema - An Evaluation
1990-09 * Ulrich Quernheim, Dieter Kreuer: Das CCITT - Signalisierungssystem Nr. 7 auf Satellitenstrecken; Simulation der Zeichengabestrecke
1990-11 H. Kuchen, R. Loogen, J.J. Moreno Navarro, M. Rodriguez Artalejo: Lazy Narrowing in a Graph Machine
1990-12 * Kai Jakobs, Josef Kaltwasser, Frank Reichert, Otto Spaniol: Der Computer fährt mit
1990-13 * Rudolf Mathar, Andreas Mann: Analyzing a Distributed Slot Assignment Protocol by Markov Chains
1990-14 A. Maassen: Compilerentwicklung in Miranda - ein Praktikum in funktionaler Programmierung (written in german)
1990-15 * Manfred Nagl, Andreas Schürr: A Specification Environment for Graph Grammars
1990-16 A. Schürr: PROGRESS: A VHL-Language Based on Graph Grammars
1990-17 * Marita Möller: Ein Ebenenmodell wissensbasierter Konsultationen - Unterstützung für Wissensakquisition und Erklärungsfähigkeit
1990-18 * Eric Kowalewski: Entwurf und Interpretation einer Sprache zur Beschreibung von Konsultationsphasen in Expertensystemen
1990-20 Y. Ortega Mallen, D. de Frutos Escrig: A Complete Proof System for Timed Observations
1990-21 * Manfred Nagl: Modelling of Software Architectures: Importance, Notions, Experiences
1990-22 H. Fassbender, H. Vogler: A Call-by-need Implementation of Syntax Directed Functional Programming
1991-01 Guenther Geiler (ed.), Fachgruppe Informatik: Jahresbericht 1990
1991-03 B. Steffen, A. Ingolfsdottir: Characteristic Formulae for Processes with Divergence
1991-04 M. Portz: A new class of cryptosystems based on interconnection networks

1991-05 H. Kuchen, G. Geiler: Distributed Applicative Arrays
1991-06 * Ludwig Staiger: Kolmogorov Complexity and Hausdorff Dimension
1991-07 * Ludwig Staiger: Syntactic Congruences for w-languages
1991-09 * Eila Kuikka: A Proposal for a Syntax-Directed Text Processing System
1991-10 K. Gladitz, H. Fassbender, H. Vogler: Compiler-based Implementation of Syntax-Directed Functional Programming
1991-11 R. Loogen, St. Winkler: Dynamic Detection of Determinism in Functional Logic Languages
1991-12 * K. Indermark, M. Rodriguez Artalejo (Eds.): Granada Workshop on the Integration of Functional and Logic Programming
1991-13 * Rolf Hager, Wolfgang Kremer: The Adaptive Priority Scheduler: A More Fair Priority Service Discipline
1991-14 * Andreas Fasbender, Wolfgang Kremer: A New Approximation Algorithm for Tandem Networks with Priority Nodes
1991-15 J. Börstler, A. Zündorf: Revisiting extensions to Modula-2 to support reusability
1991-16 J. Börstler, Th. Janning: Bridging the gap between Requirements Analysis and Design
1991-17 A. Zündorf, A. Schürr: Nondeterministic Control Structures for Graph Rewriting Systems
1991-18 * Matthias Jarke, John Mylopoulos, Joachim W. Schmidt, Yannis Vassiliou: DAIDA: An Environment for Evolving Information Systems
1991-19 M. Jeusfeld, M. Jarke: From Relational to Object-Oriented Integrity Simplification
1991-20 G. Hogen, A. Kindler, R. Loogen: Automatic Parallelization of Lazy Functional Programs
1991-21 * Prof. Dr. rer. nat. Otto Spaniol: ODP (Open Distributed Processing): Yet another Viewpoint
1991-22 H. Kuchen, F. Lücking, H. Stoltze: The Topology Description Language TDL
1991-23 S. Graf, B. Steffen: Compositional Minimization of Finite State Systems
1991-24 R. Cleaveland, J. Parrow, B. Steffen: The Concurrency Workbench: A Semantics Based Tool for the Verification of Concurrent Systems
1991-25 * Rudolf Mathar, Jürgen Mattfeldt: Optimal Transmission Ranges for Mobile Communication in Linear Multihop Packet Radio Networks
1991-26 M. Jeusfeld, M. Staudt: Query Optimization in Deductive Object Bases
1991-27 J. Knoop, B. Steffen: The Interprocedural Coincidence Theorem
1991-28 J. Knoop, B. Steffen: Unifying Strength Reduction and Semantic Code Motion
1991-30 T. Margaria: First-Order theories for the verification of complex FSMs
1991-31 B. Steffen: Generating Data Flow Analysis Algorithms from Modal Specifications
1992-01 Stefan Eherer (ed.), Fachgruppe Informatik: Jahresbericht 1991
1992-02 * Bernhard Westfechtel: Basismechanismen zur Datenverwaltung in strukturbezogenen Hypertextsystemen
1992-04 S. A. Smolka, B. Steffen: Priority as Extremal Probability
1992-05 * Matthias Jarke, Carlos Maltzahn, Thomas Rose: Sharing Processes: Team Coordination in Design Repositories

1992-06 O. Burkart, B. Steffen: Model Checking for Context-Free Processes

1992-07 * Matthias Jarke, Klaus Pohl: Information Systems Quality and Quality Information Systems

1992-08 * Rudolf Mathar, Jürgen Mattfeldt: Analyzing Routing Strategy NFP in Multihop Packet Radio Networks on a Line

1992-09 * Alfons Kemper, Guido Moerkotte: Grundlagen objektorientierter Datenbanksysteme

1992-10 Matthias Jarke, Manfred Jeusfeld, Andreas Miethsam, Michael Gocek: Towards a logic-based reconstruction of software configuration management

1992-11 Werner Hans: A Complete Indexing Scheme for WAM-based Abstract Machines

1992-12 W. Hans, R. Loogen, St. Winkler: On the Interaction of Lazy Evaluation and Backtracking

1992-13 * Matthias Jarke, Thomas Rose: Specification Management with CAD

1992-14 Th. Noll, H. Vogler: Top-down Parsing with Simultaneous Evaluation on Noncircular Attribute Grammars

1992-15 A. Schuerr, B. Westfechtel: Graphgrammatiken und Graphersetzungssysteme(written in german)

1992-16 * Graduiertenkolleg Informatik und Technik (Hrsg.): Forschungsprojekte des Graduiertenkollegs Informatik und Technik

1992-17 M. Jarke (ed.): ConceptBase V3.1 User Manual

1992-18 * Clarence A. Ellis, Matthias Jarke (Eds.): Distributed Cooperation in Integrated Information Systems - Proceedings of the Third International Workshop on Intelligent and Cooperative Information Systems

1992-19-00 H. Kuchen, R. Loogen (eds.): Proceedings of the 4th Int. Workshop on the Parallel Implementation of Functional Languages

1992-19-01 G. Hogen, R. Loogen: PASTEL - A Parallel Stack-Based Implementation of Eager Functional Programs with Lazy Data Structures (Extended Abstract)

1992-19-02 H. Kuchen, K. Gladitz: Implementing Bags on a Shared Memory MIMD-Machine

1992-19-03 C. Rathsack, S.B. Scholz: LISA - A Lazy Interpreter for a Full-Fledged Lambda-Calculus

1992-19-04 T.A. Bratvold: Determining Useful Parallelism in Higher Order Functions

1992-19-05 S. Kahrs: Polymorphic Type Checking by Interpretation of Code

1992-19-06 M. Chakravarty, M. Köhler: Equational Constraints, Residuation, and the Parallel JUMP-Machine

1992-19-07 J. Seward: Polymorphic Strictness Analysis using Frontiers (Draft Version)

1992-19-08 D. Gärtner, A. Kimms, W. Kluge: pi-Red^+ - A Compiling Graph-Reduction System for a Full Fledged Lambda-Calculus

1992-19-09 D. Howe, G. Burn: Experiments with strict STG code

1992-19-10 J. Glauert: Parallel Implementation of Functional Languages Using Small Processes

1992-19-11 M. Joy, T. Axford: A Parallel Graph Reduction Machine

1992-19-12 A. Bennett, P. Kelly: Simulation of Multicache Parallel Reduction

1992-19-13 K. Langendoen, D.J. Agterkamp: Cache Behaviour of Lazy Functional Programs (Working Paper)
1992-19-14 K. Hammond, S. Peyton Jones: Profiling scheduling strategies on the GRIP parallel reducer
1992-19-15 S. Mintchev: Using Strictness Information in the STG-machine
1992-19-16 D. Rushall: An Attribute Grammar Evaluator in Haskell
1992-19-17 J. Wild, H. Glaser, P. Hartel: Statistics on storage management in a lazy functional language implementation
1992-19-18 W.S. Martins: Parallel Implementations of Functional Languages
1992-19-19 D. Lester: Distributed Garbage Collection of Cyclic Structures (Draft version)
1992-19-20 J.C. Glas, R.F.H. Hofman, W.G. Vree: Parallelization of Branch-and-Bound Algorithms in a Functional Programming Environment
1992-19-21 S. Hwang, D. Rushall: The nu-STG machine: a parallelized Spineless Tagless Graph Reduction Machine in a distributed memory architecture (Draft version)
1992-19-22 G. Burn, D. Le Metayer: Cps-Translation and the Correctness of Optimising Compilers
1992-19-23 S.L. Peyton Jones, P. Wadler: Imperative functional programming (Brief summary)
1992-19-24 W. Damm, F. Liu, Th. Peikenkamp: Evaluation and Parallelization of Functions in Functional + Logic Languages (abstract)
1992-19-25 M. Kesseler: Communication Issues Regarding Parallel Functional Graph Rewriting
1992-19-26 Th. Peikenkamp: Charakterizing and representing neededness in functional loginc languages (abstract)
1992-19-27 H. Doerr: Monitoring with Graph-Grammars as formal operational Models
1992-19-28 J. van Groningen: Some implementation aspects of Concurrent Clean on distributed memory architectures
1992-19-29 G. Ostheimer: Load Bounding for Implicit Parallelism (abstract)
1992-20 H. Kuchen, F.J. Lopez Fraguas, J.J. Moreno Navarro, M. Rodriguez Artalejo: Implementing Disequality in a Lazy Functional Logic Language
1992-21 H. Kuchen, F.J. Lopez Fraguas: Result Directed Computing in a Functional Logic Language
1992-22 H. Kuchen, J.J. Moreno Navarro, M.V. Hermenegildo: Independent AND-Parallel Narrowing
1992-23 T. Margaria, B. Steffen: Distinguishing Formulas for Free
1992-24 K. Pohl: The Three Dimensions of Requirements Engineering
1992-25 * R. Stainov: A Dynamic Configuration Facility for Multimedia Communications
1992-26 * Michael von der Beeck: Integration of Structured Analysis and Timed Statecharts for Real-Time and Concurrency Specification
1992-27 W. Hans, St. Winkler: Aliasing and Groundness Analysis of Logic Programs through Abstract Interpretation and its Safety
1992-28 * Gerhard Steinke, Matthias Jarke: Support for Security Modeling in Information Systems Design
1992-29 B. Schinzel: Warum Frauenforschung in Naturwissenschaft und Technik

1992-30 A. Kemper, G. Moerkotte, K. Peithner: Object-Orientation Axiomatised by Dynamic Logic

1992-32 * Bernd Heinrichs, Kai Jakobs: Timer Handling in High-Performance Transport Systems

1992-33 * B. Heinrichs, K. Jakobs, K. Lenßen, W. Reinhardt, A. Spinner: Euro-Bridge: Communication Services for Multimedia Applications

1992-34 C. Gerlhof, A. Kemper, Ch. Kilger, G. Moerkotte: Partition-Based Clustering in Object Bases: From Theory to Practice

1992-35 J. Börstler: Feature-Oriented Classification and Reuse in IPSEN

1992-36 M. Jarke, J. Bubenko, C. Rolland, A. Sutcliffe, Y. Vassiliou: Theories Underlying Requirements Engineering: An Overview of NATURE at Genesis

1992-37 * K. Pohl, M. Jarke: Quality Information Systems: Repository Support for Evolving Process Models

1992-38 A. Zuendorf: Implementation of the imperative / rule based language PROGRES

1992-39 P. Koch: Intelligentes Backtracking bei der Auswertung funktional-logischer Programme

1992-40 * Rudolf Mathar, Jürgen Mattfeldt: Channel Assignment in Cellular Radio Networks

1992-41 * Gerhard Friedrich, Wolfgang Neidl: Constructive Utility in Model-Based Diagnosis Repair Systems

1992-42 * P. S. Chen, R. Hennicker, M. Jarke: On the Retrieval of Reusable Software Components

1992-43 W. Hans, St.Winkler: Abstract Interpretation of Functional Logic Languages

1992-44 N. Kiesel, A. Schuerr, B. Westfechtel: Design and Evaluation of GRAS, a Graph-Oriented Database System for Engineering Applications

1993-01 * Fachgruppe Informatik: Jahresbericht 1992

1993-02 * Patrick Shicheng Chen: On Inference Rules of Logic-Based Information Retrieval Systems

1993-03 G. Hogen, R. Loogen: A New Stack Technique for the Management of Runtime Structures in Distributed Environments

1993-05 A. Zündorf: A Heuristic for the Subgraph Isomorphism Problem in Executing PROGRES

1993-06 A. Kemper, D. Kossmann: Adaptable Pointer Swizzling Strategies in Object Bases: Design, Realization, and Quantitative Analysis

1993-07 * Graduiertenkolleg Informatik und Technik (Hrsg.): Graduiertenkolleg Informatik und Technik

1993-08 * Matthias Berger: k-Coloring Vertices using a Neural Network with Convergence to Valid Solutions

1993-09 M. Buchheit, M. Jeusfeld, W. Nutt, M. Staudt: Subsumption between Queries to Object-Oriented Databases

1993-10 O. Burkart, B. Steffen: Pushdown Processes: Parallel Composition and Model Checking

1993-11 * R. Große-Wienker, O. Hermanns, D. Menzenbach, A. Pollacks, S. Repetzki, J. Schwartz, K. Sonnenschein, B. Westfechtel: Das SUKITS-Projekt: A-posteriori-Integration heterogener CIM-Anwendungssysteme 1993-12 * Rudolf Mathar, Jürgen Mattfeldt: On the Distribution of Cumulated Interference Power in Rayleigh Fading Channels

1993-13 O. Maler, L. Staiger: On Syntactic Congruences for omega-languages

1993-14 M. Jarke, St. Eherer, R. Gallersdoerfer, M. Jeusfeld, M. Staudt: ConceptBase - A Deductive Object Base Manager

1993-15 M. Staudt, H.W. Nissen, M.A. Jeusfeld: Query by Class, Rule and Concept

1993-16 * M. Jarke, K. Pohl, St. Jacobs et al.: Requirements Engineering: An Integrated View of Representation Process and Domain

1993-17 * M. Jarke, K. Pohl: Establishing Vision in Context: Towards a Model of Requirements Processes

1993-18 W. Hans, H. Kuchen, St. Winkler: Full Indexing for Lazy Narrowing

1993-19 W. Hans, J.J. Ruz, F. Saenz, St. Winkler: A VHDL Specification of a Shared Memory Parallel Machine for Babel

1993-20 * K. Finke, M. Jarke, P. Szczurko, R. Soltysiak: Quality Management for Expert Systems in Process Control

1993-21 M. Jarke, M.A. Jeusfeld, P. Szczurko: Three Aspects of Intelligent Cooperation in the Quality Cycle

1994-01 Margit Generet, Sven Martin (eds.), Fachgruppe Informatik: Jahresbericht 1993

1994-02 M. Lefering: Development of Incremental Integration Tools Using Formal Specifications

1994-03 * P. Constantopoulos, M. Jarke, J. Mylopoulos, Y. Vassiliou: The Software Information Base: A Server for Reuse

1994-04 * Rolf Hager, Rudolf Mathar, Jürgen Mattfeldt: Intelligent Cruise Control and Reliable Communication of Mobile Stations

1994-05 * Rolf Hager, Peter Hermesmann, Michael Portz: Feasibility of Authentication Procedures within Advanced Transport Telematics

1994-06 * Claudia Popien, Bernd Meyer, Axel Kuepper: A Formal Approach to Service Import in ODP Trader Federations

1994-07 P. Peters, P. Szczurko: Integrating Models of Quality Management Methods by an Object-Oriented Repository

1994-08 * Manfred Nagl, Bernhard Westfechtel: A Universal Component for the Administration in Distributed and Integrated Development Environments

1994-09 * Patrick Horster, Holger Petersen: Signatur- und Authentifikationsverfahren auf der Basis des diskreten Logarithmusproblems

1994-11 A. Schürr: PROGRES, A Visual Language and Environment for PROgramming with Graph REwrite Systems

1994-12 A. Schürr: Specification of Graph Translators with Triple Graph Grammars

1994-13 A. Schürr: Logic Based Programmed Structure Rewriting Systems

1994-14 L. Staiger: Codes, Simplifying Words, and Open Set Condition

1994-15 * Bernhard Westfechtel: A Graph-Based System for Managing Configurations of Engineering Design Documents

1994-16 P. Klein: Designing Software with Modula-3

1994-17 I. Litovsky, L. Staiger: Finite acceptance of infinite words

1994-18 G. Hogen, R. Loogen: Parallel Functional Implementations: Graphbased vs. Stackbased Reduction

1994-19 M. Jeusfeld, U. Johnen: An Executable Meta Model for Re-Engineering of Database Schemas

1994-20 * R. Gallersdörfer, M. Jarke, K. Klabunde: Intelligent Networks as a Data Intensive Application (INDIA)

1994-21 M. Mohnen: Proving the Correctness of the Static Link Technique Using Evolving Algebras

1994-22 H. Fernau, L. Staiger: Valuations and Unambiguity of Languages, with Applications to Fractal Geometry

1994-24 * M. Jarke, K. Pohl, R. Dömges, St. Jacobs, H. W. Nissen: Requirements Information Management: The NATURE Approach

1994-25 * M. Jarke, K. Pohl, C. Rolland, J.-R. Schmitt: Experience-Based Method Evaluation and Improvement: A Process Modeling Approach

1994-26 * St. Jacobs, St. Kethers: Improving Communication and Decision Making within Quality Function Deployment

1994-27 * M. Jarke, H. W. Nissen, K. Pohl: Tool Integration in Evolving Information Systems Environments

1994-28 O. Burkart, D. Caucal, B. Steffen: An Elementary Bisimulation Decision Procedure for Arbitrary Context-Free Processes

1995-01 * Fachgruppe Informatik: Jahresbericht 1994

1995-02 Andy Schürr, Andreas J. Winter, Albert Zündorf: Graph Grammar Engineering with PROGRES

1995-03 Ludwig Staiger: A Tight Upper Bound on Kolmogorov Complexity by Hausdorff Dimension and Uniformly Optimal Prediction

1995-04 Birgitta König-Ries, Sven Helmer, Guido Moerkotte: An experimental study on the complexity of left-deep join ordering problems for cyclic queries

1995-05 Sophie Cluet, Guido Moerkotte: Efficient Evaluation of Aggregates on Bulk Types

1995-06 Sophie Cluet, Guido Moerkotte: Nested Queries in Object Bases

1995-07 Sophie Cluet, Guido Moerkotte: Query Optimization Techniques Exploiting Class Hierarchies

1995-08 Markus Mohnen: Efficient Compile-Time Garbage Collection for Arbitrary Data Structures

1995-09 Markus Mohnen: Functional Specification of Imperative Programs: An Alternative Point of View of Functional Languages

1995-10 Rainer Gallersdörfer, Matthias Nicola: Improving Performance in Replicated Databases through Relaxed Coherency

1995-11 * M.Staudt, K.von Thadden: Subsumption Checking in Knowledge Bases

1995-12 * G.V.Zemanek, H.W.Nissen, H.Hubert, M.Jarke: Requirements Analysis from Multiple Perspectives: Experiences with Conceptual Modeling Technology

1995-13 * M.Staudt, M.Jarke: Incremental Maintenance of Externally Materialized Views

1995-14 * P.Peters, P.Szczurko, M.Jeusfeld: Oriented Information Management: Conceptual Models at Work

1995-15 * Matthias Jarke, Sudha Ram (Hrsg.): WITS 95 Proceedings of the 5th Annual Workshop on Information Technologies and Systems
1995-16 * W.Hans, St.Winkler, F.Saenz: Distributed Execution in Functional Logic Programming
1996-01 * Jahresbericht 1995
1996-02 Michael Hanus, Christian Prehofer: Higher-Order Narrowing with Definitional Trees
1996-03 * W.Scheufele, G.Moerkotte: Optimal Ordering of Selections and Joins in Acyclic Queries with Expensive Predicates
1996-04 Klaus Pohl: PRO-ART: Enabling Requirements Pre-Traceability
1996-05 Klaus Pohl: Requirements Engineering: An Overview
1996-06 * M.Jarke, W.Marquardt: Design and Evaluation of Computer–Aided Process Modelling Tools
1996-07 Olaf Chitil: The Sigma-Semantics: A Comprehensive Semantics for Functional Programs
1996-08 * S.Sripada: On Entropy and the Limitations of the Second Law of Thermodynamics
1996-09 Michael Hanus (Ed.): Proceedings of the Poster Session of ALP96 - Fifth International Conference on Algebraic and Logic Programming
1996-09-0 Michael Hanus (Ed.): Proceedings of the Poster Session of ALP 96 - Fifth International Conference on Algebraic and Logic Programming: Introduction and table of contents
1996-09-1 Ilies Alouini: An Implementation of Conditional Concurrent Rewriting on Distributed Memory Machines
1996-09-2 Olivier Danvy, Karoline Malmkjær: On the Idempotence of the CPS Transformation
1996-09-3 Víctor M. Gulías, José L. Freire: Concurrent Programming in Haskell
1996-09-4 Sébastien Limet, Pierre Réty: On Decidability of Unifiability Modulo Rewrite Systems
1996-09-5 Alexandre Tessier: Declarative Debugging in Constraint Logic Programming
1996-10 Reidar Conradi, Bernhard Westfechtel: Version Models for Software Configuration Management
1996-11 * C.Weise, D.Lenzkes: A Fast Decision Algorithm for Timed Refinement
1996-12 * R.Dömges, K.Pohl, M.Jarke, B.Lohmann, W.Marquardt: PRO-ART/CE* — An Environment for Managing the Evolution of Chemical Process Simulation Models
1996-13 * K.Pohl, R.Klamma, K.Weidenhaupt, R.Dömges, P.Haumer, M.Jarke: A Framework for Process-Integrated Tools
1996-14 * R.Gallersdörfer, K.Klabunde, A.Stolz, M.Eßmajor: INDIA — Intelligent Networks as a Data Intensive Application, Final Project Report, June 1996
1996-15 * H.Schimpe, M.Staudt: VAREX: An Environment for Validating and Refining Rule Bases
1996-16 * M.Jarke, M.Gebhardt, S.Jacobs, H.Nissen: Conflict Analysis Across Heterogeneous Viewpoints: Formalization and Visualization
1996-17 Manfred A. Jeusfeld, Tung X. Bui: Decision Support Components on the Internet

1996-18 Manfred A. Jeusfeld, Mike Papazoglou: Information Brokering: Design, Search and Transformation

1996-19 * P.Peters, M.Jarke: Simulating the impact of information flows in networked organizations

1996-20 Matthias Jarke, Peter Peters, Manfred A. Jeusfeld: Model-driven planning and design of cooperative information systems

1996-21 * G.de Michelis, E.Dubois, M.Jarke, F.Matthes, J.Mylopoulos, K.Pohl, J.Schmidt, C.Woo, E.Yu: Cooperative information systems: a manifesto

1996-22 * S.Jacobs, M.Gebhardt, S.Kethers, W.Rzasa: Filling HTML forms simultaneously: CoWeb architecture and functionality

1996-23 * M.Gebhardt, S.Jacobs: Conflict Management in Design

1997-01 Michael Hanus, Frank Zartmann (eds.): Jahresbericht 1996

1997-02 Johannes Faassen: Using full parallel Boltzmann Machines for Optimization

1997-03 Andreas Winter, Andy Schürr: Modules and Updatable Graph Views for PROgrammed Graph REwriting Systems

1997-04 Markus Mohnen, Stefan Tobies: Implementing Context Patterns in the Glasgow Haskell Compiler

1997-05 * S.Gruner: Schemakorrespondenzaxiome unterstützen die paargrammatische Spezifikation inkrementeller Integrationswerkzeuge

1997-06 Matthias Nicola, Matthias Jarke: Design and Evaluation of Wireless Health Care Information Systems in Developing Countries

1997-07 Petra Hofstedt: Taskparallele Skelette für irregulär strukturierte Probleme in deklarativen Sprachen

1997-08 Dorothea Blostein, Andy Schürr: Computing with Graphs and Graph Rewriting

1997-09 Carl-Arndt Krapp, Bernhard Westfechtel: Feedback Handling in Dynamic Task Nets

1997-10 Matthias Nicola, Matthias Jarke: Integrating Replication and Communication in Performance Models of Distributed Databases

1997-11 * R. Klamma, P. Peters, M. Jarke: Workflow Support for Failure Management in Federated Organizations

1997-13 Markus Mohnen: Optimising the Memory Management of Higher-Order Functional Programs

1997-14 Roland Baumann: Client/Server Distribution in a Structure-Oriented Database Management System

1997-15 George Botorog: High-Level Parallel Programming and the Efficient Implementation of Numerical Algorithms

1998-01 * Fachgruppe Informatik: Jahresbericht 1997

1998-02 Stefan Gruner, Manfred Nagel, Andy Schürr: Fine-grained and Structure-Oriented Document Integration Tools are Needed for Development Processes

1998-03 Stefan Gruner: Einige Anmerkungen zur graphgrammatischen Spezifikation von Integrationswerkzeugen nach Westfechtel, Janning, Lefering und Schürr

1998-04 * O. Kubitz: Mobile Robots in Dynamic Environments

1998-05 Martin Leucker, Stephan Tobies: Truth - A Verification Platform for Distributed Systems

1998-06 * Matthias Oliver Berger: DECT in the Factory of the Future
1998-07 M. Arnold, M. Erdmann, M. Glinz, P. Haumer, R. Knoll, B. Paech, K. Pohl, J. Ryser, R. Studer, K. Weidenhaupt: Survey on the Scenario Use in Twelve Selected Industrial Projects
1998-09 * Th. Lehmann: Geometrische Ausrichtung medizinischer Bilder am Beispiel intraoraler Radiographien
1998-10 * M. Nicola, M. Jarke: Performance Modeling of Distributed and Replicated Databases
1998-11 * Ansgar Schleicher, Bernhard Westfechtel, Dirk Jäger: Modeling Dynamic Software Processes in UML
1998-12 * W. Appelt, M. Jarke: Interoperable Tools for Cooperation Support using the World Wide Web
1998-13 Klaus Indermark: Semantik rekursiver Funktionsdefinitionen mit Striktheitsinformation
1999-01 * Jahresbericht 1998
1999-02 * F. Huch: Verifcation of Erlang Programs using Abstract Interpretation and Model Checking — Extended Version
1999-03 * R. Gallersdörfer, M. Jarke, M. Nicola: The ADR Replication Manager
1999-04 María Alpuente, Michael Hanus, Salvador Lucas, Germán Vidal: Specialization of Functional Logic Programs Based on Needed Narrowing
1999-05 * W. Thomas (Ed.): DLT 99 - Developments in Language Theory Fourth International Conference
1999-06 * Kai Jakobs, Klaus-Dieter Kleefeld: Informationssysteme für die angewandte historische Geographie
1999-07 Thomas Wilke: CTL+ is exponentially more succinct than CTL
1999-08 Oliver Matz: Dot-Depth and Monadic Quantifier Alternation over Pictures
2000-01 * Jahresbericht 1999
2000-02 Jens Vöge, Marcin Jurdzinski A Discrete Strategy Improvement Algorithm for Solving Parity Games
2000-03 D. Jäger, A. Schleicher, B. Westfechtel: UPGRADE: A Framework for Building Graph-Based Software Engineering Tools
2000-04 Andreas Becks, Stefan Sklorz, Matthias Jarke: Exploring the Semantic Structure of Technical Document Collections: A Cooperative Systems Approach
2000-05 Mareike Schoop: Cooperative Document Management
2000-06 Mareike Schoop, Christoph Quix (eds.): Proceedings of the Fifth International Workshop on the Language-Action Perspective on Communication Modelling
2000-07 * Markus Mohnen, Pieter Koopman (Eds.): Proceedings of the 12th International Workshop of Functional Languages
2000-08 Thomas Arts, Thomas Noll: Verifying Generic Erlang Client-Server Implementations
2001-01 * Jahresbericht 2000
2001-02 Benedikt Bollig, Martin Leucker: Deciding LTL over Mazurkiewicz Traces
2001-03 Thierry Cachat: The power of one-letter rational languages

2001-04 Benedikt Bollig, Martin Leucker, Michael Weber: Local Parallel Model Checking for the Alternation Free mu-Calculus

2001-05 Benedikt Bollig, Martin Leucker, Thomas Noll: Regular MSC Languages

2001-06 Achim Blumensath: Prefix-Recognisable Graphs and Monadic Second-Order Logic

2001-07 Martin Grohe, Stefan Wöhrle: An Existential Locality Theorem

2001-08 Mareike Schoop, James Taylor (eds.): Proceedings of the Sixth International Workshop on the Language-Action Perspective on Communication Modelling

2001-09 Thomas Arts, Jürgen Giesl: A collection of examples for termination of term rewriting using dependency pairs

2001-10 Achim Blumensath: Axiomatising Tree-interpretable Structures

2001-11 Klaus Indermark, Thomas Noll (eds.): Kolloquium Programmiersprachen und Grundlagen der Programmierung

2002-01 * Jahresbericht 2001

2002-02 Jürgen Giesl, Aart Middeldorp: Transformation Techniques for Context-Sensitive Rewrite Systems

2002-03 Benedikt Bollig, Martin Leucker, Thomas Noll: Generalised Regular MSC Languages

2002-04 Jürgen Giesl, Aart Middeldorp: Innermost Termination of Context-Sensitive Rewriting

2002-05 Horst Lichter, Thomas von der Maßen, Thomas Weiler: Modelling Requirements and Architectures for Software Product Lines

2002-06 Henry N. Adorna: 3-Party Message Complexity is Better than 2-Party Ones for Proving Lower Bounds on the Size of Minimal Nondeterministic Finite Automata

2002-07 Jörg Dahmen: Invariant Image Object Recognition using Gaussian Mixture Densities

2002-08 Markus Mohnen: An Open Framework for Data-Flow Analysis in Java

2002-09 Markus Mohnen: Interfaces with Default Implementations in Java

2002-10 Martin Leucker: Logics for Mazurkiewicz traces

2002-11 Jürgen Giesl, Hans Zantema: Liveness in Rewriting

2003-01 * Jahresbericht 2002

2003-02 Jürgen Giesl, René Thiemann: Size-Change Termination for Term Rewriting

2003-03 Jürgen Giesl, Deepak Kapur: Deciding Inductive Validity of Equations

2003-04 Jürgen Giesl, René Thiemann, Peter Schneider-Kamp, Stephan Falke: Improving Dependency Pairs

2003-05 Christof Löding, Philipp Rohde: Solving the Sabotage Game is PSPACE-hard

2003-06 Franz Josef Och: Statistical Machine Translation: From Single-Word Models to Alignment Templates

2003-07 Horst Lichter, Thomas von der Maßen, Alexander Nyßen, Thomas Weiler: Vergleich von Ansätzen zur Feature Modellierung bei der Softwareproduktlinienentwicklung

2003-08 Jürgen Giesl, René Thiemann, Peter Schneider-Kamp, Stephan Falke: Mechanizing Dependency Pairs

2004-01 * Fachgruppe Informatik: Jahresbericht 2003

2004-02 Benedikt Bollig, Martin Leucker: Message-Passing Automata are expressively equivalent to EMSO logic
2004-03 Delia Kesner, Femke van Raamsdonk, Joe Wells (eds.): HOR 2004 – 2nd International Workshop on Higher-Order Rewriting
2004-04 Slim Abdennadher, Christophe Ringeissen (eds.): RULE 04 – Fifth International Workshop on Rule-Based Programming
2004-05 Herbert Kuchen (ed.): WFLP 04 – 13th International Workshop on Functional and (Constraint) Logic Programming
2004-06 Sergio Antoy, Yoshihito Toyama (eds.): WRS 04 – 4th International Workshop on Reduction Strategies in Rewriting and Programming
2004-07 Michael Codish, Aart Middeldorp (eds.): WST 04 – 7th International Workshop on Termination
2004-08 Klaus Indermark, Thomas Noll: Algebraic Correctness Proofs for Compiling Recursive Function Definitions with Strictness Information
2004-09 Joachim Kneis, Daniel Mölle, Stefan Richter, Peter Rossmanith: Parameterized Power Domination Complexity
2004-10 Zinaida Benenson, Felix C. Gärtner, Dogan Kesdogan: Secure Multi-Party Computation with Security Modules
2005-01 * Fachgruppe Informatik: Jahresbericht 2004
2005-02 Maximillian Dornseif, Felix C. Gärtner, Thorsten Holz, Martin Mink: An Offensive Approach to Teaching Information Security: "Aachen Summer School Applied IT Security"
2005-03 Jürgen Giesl, René Thiemann, Peter Schneider-Kamp: Proving and Disproving Termination of Higher-Order Functions
2005-04 Daniel Mölle, Stefan Richter, Peter Rossmanith: A Faster Algorithm for the Steiner Tree Problem
2005-05 Fabien Pouget, Thorsten Holz: A Pointillist Approach for Comparing Honeypots
2005-06 Simon Fischer, Berthold Vöcking: Adaptive Routing with Stale Information
2005-07 Felix C. Freiling, Thorsten Holz, Georg Wicherski: Botnet Tracking: Exploring a Root-Cause Methodology to Prevent Distributed Denial-of-Service Attacks
2005-08 Joachim Kneis, Peter Rossmanith: A New Satisfiability Algorithm With Applications To Max-Cut
2005-09 Klaus Kursawe, Felix C. Freiling: Byzantine Fault Tolerance on General Hybrid Adversary Structures
2005-10 Benedikt Bollig: Automata and Logics for Message Sequence Charts
2005-11 Simon Fischer, Berthold Vöcking: A Counterexample to the Fully Mixed Nash Equilibrium Conjecture
2005-12 Neeraj Mittal, Felix Freiling, S. Venkatesan, Lucia Draque Penso: Efficient Reductions for Wait-Free Termination Detection in Faulty Distributed Systems
2005-13 Carole Delporte-Gallet, Hugues Fauconnier, Felix C. Freiling: Revisiting Failure Detection and Consensus in Omission Failure Environments
2005-14 Felix C. Freiling, Sukumar Ghosh: Code Stabilization
2005-15 Uwe Naumann: The Complexity of Derivative Computation

2005-16 Uwe Naumann: Syntax-Directed Derivative Code (Part I: Tangent-Linear Code)

2005-17 Uwe Naumann: Syntax-directed Derivative Code (Part II: Intraprocedural Adjoint Code)

2005-18 Thomas von der Maßen, Klaus Müller, John MacGregor, Eva Geisberger, Jörg Dörr, Frank Houdek, Harbhajan Singh, Holger Wußmann, Hans-Veit Bacher, Barbara Paech: Einsatz von Features im Software-Entwicklungsprozess - Abschlußbericht des GI-Arbeitskreises "Features"

2005-19 Uwe Naumann, Andre Vehreschild: Tangent-Linear Code by Augmented LL-Parsers

2005-20 Felix C. Freiling, Martin Mink: Bericht über den Workshop zur Ausbildung im Bereich IT-Sicherheit Hochschulausbildung, berufliche Weiterbildung, Zertifizierung von Ausbildungsangeboten am 11. und 12. August 2005 in Köln organisiert von RWTH Aachen in Kooperation mit BITKOM, BSI, DLR und Gesellschaft fuer Informatik (GI) e.V.

2005-21 Thomas Noll, Stefan Rieger: Optimization of Straight-Line Code Revisited

2005-22 Felix Freiling, Maurice Herlihy, Lucia Draque Penso: Optimal Randomized Fair Exchange with Secret Shared Coins

2005-23 Heiner Ackermann, Alantha Newman, Heiko Röglin, Berthold Vöcking: Decision Making Based on Approximate and Smoothed Pareto Curves

2005-24 Alexander Becher, Zinaida Benenson, Maximillian Dornseif: Tampering with Motes: Real-World Physical Attacks on Wireless Sensor Networks

2006-01 * Fachgruppe Informatik: Jahresbericht 2005

2006-02 Michael Weber: Parallel Algorithms for Verification of Large Systems

2006-03 Michael Maier, Uwe Naumann: Intraprocedural Adjoint Code Generated by the Differentiation-Enabled NAGWare Fortran Compiler

2006-04 Ebadollah Varnik, Uwe Naumann, Andrew Lyons: Toward Low Static Memory Jacobian Accumulation

2006-05 Uwe Naumann, Jean Utke, Patrick Heimbach, Chris Hill, Derya Ozyurt, Carl Wunsch, Mike Fagan, Nathan Tallent, Michelle Strout: Adjoint Code by Source Transformation with OpenAD/F

2006-06 Joachim Kneis, Daniel Mölle, Stefan Richter, Peter Rossmanith: Divide-and-Color

2006-07 Thomas Colcombet, Christof Löding: Transforming structures by set interpretations

2006-08 Uwe Naumann, Yuxiao Hu: Optimal Vertex Elimination in Single-Expression-Use Graphs

2006-09 Tingting Han, Joost-Pieter Katoen: Counterexamples in Probabilistic Model Checking

2006-10 Mesut Günes, Alexander Zimmermann, Martin Wenig, Jan Ritzerfeld, Ulrich Meis: From Simulations to Testbeds - Architecture of the Hybrid MCG-Mesh Testbed

2006-11 Bastian Schlich, Michael Rohrbach, Michael Weber, Stefan Kowalewski: Model Checking Software for Microcontrollers

2006-12 Benedikt Bollig, Joost-Pieter Katoen, Carsten Kern, Martin Leucker: Replaying Play in and Play out: Synthesis of Design Models from Scenarios by Learning

2006-13 Wong Karianto, Christof Löding: Unranked Tree Automata with Sibling Equalities and Disequalities

2006-14 Danilo Beuche, Andreas Birk, Heinrich Dreier, Andreas Fleischmann, Heidi Galle, Gerald Heller, Dirk Janzen, Isabel John, Ramin Tavakoli Kolagari, Thomas von der Maßen, Andreas Wolfram: Report of the GI Work Group "Requirements Management Tools for Product Line Engineering"

2006-15 Sebastian Ullrich, Jakob T. Valvoda, Torsten Kuhlen: Utilizing optical sensors from mice for new input devices

2006-16 Rafael Ballagas, Jan Borchers: Selexels: a Conceptual Framework for Pointing Devices with Low Expressiveness

2006-17 Eric Lee, Henning Kiel, Jan Borchers: Scrolling Through Time: Improving Interfaces for Searching and Navigating Continuous Audio Timelines

2007-01 * Fachgruppe Informatik: Jahresbericht 2006

2007-02 Carsten Fuhs, Jürgen Giesl, Aart Middeldorp, Peter Schneider-Kamp, René Thiemann, and Harald Zankl: SAT Solving for Termination Analysis with Polynomial Interpretations

2007-03 Jürgen Giesl, René Thiemann, Stephan Swiderski, and Peter Schneider-Kamp: Proving Termination by Bounded Increase

2007-04 Jan Buchholz, Eric Lee, Jonathan Klein, and Jan Borchers: coJIVE: A System to Support Collaborative Jazz Improvisation

2007-05 Uwe Naumann: On Optimal DAG Reversal

2007-06 Joost-Pieter Katoen, Thomas Noll, and Stefan Rieger: Verifying Concurrent List-Manipulating Programs by LTL Model Checking

2007-07 Alexander Nyßen, Horst Lichter: MeDUSA - MethoD for UML2-based Design of Embedded Software Applications

2007-08 Falk Salewski and Stefan Kowalewski: Achieving Highly Reliable Embedded Software: An empirical evaluation of different approaches

2007-09 Tina Kraußer, Heiko Mantel, and Henning Sudbrock: A Probabilistic Justification of the Combining Calculus under the Uniform Scheduler Assumption

2007-10 Martin Neuhäußer, Joost-Pieter Katoen: Bisimulation and Logical Preservation for Continuous-Time Markov Decision Processes

2007-11 Klaus Wehrle (editor): 6. Fachgespräch Sensornetzwerke

2007-12 Uwe Naumann: An L-Attributed Grammar for Adjoint Code

2007-13 Uwe Naumann, Michael Maier, Jan Riehme, and Bruce Christianson: Second-Order Adjoints by Source Code Manipulation of Numerical Programs

2007-14 Jean Utke, Uwe Naumann, Mike Fagan, Nathan Tallent, Michelle Strout, Patrick Heimbach, Chris Hill, and Carl Wunsch: OpenAD/F: A Modular, Open-Source Tool for Automatic Differentiation of Fortran Codes

2007-15 Volker Stolz: Temporal assertions for sequential and concurrent programs

2007-16 Sadeq Ali Makram, Mesut Güneç, Martin Wenig, Alexander Zimmermann: Adaptive Channel Assignment to Support QoS and Load Balancing for Wireless Mesh Networks

2007-17 René Thiemann: The DP Framework for Proving Termination of Term Rewriting

2007-18 Uwe Naumann: Call Tree Reversal is NP-Complete

2007-19 Jan Riehme, Andrea Walther, Jörg Stiller, Uwe Naumann: Adjoints for Time-Dependent Optimal Control

2007-20 Joost-Pieter Katoen, Daniel Klink, Martin Leucker, and Verena Wolf: Three-Valued Abstraction for Probabilistic Systems

2007-21 Tingting Han, Joost-Pieter Katoen, and Alexandru Mereacre: Compositional Modeling and Minimization of Time-Inhomogeneous Markov Chains

2007-22 Heiner Ackermann, Paul W. Goldberg, Vahab S. Mirrokni, Heiko Röglin, and Berthold Vöcking: Uncoordinated Two-Sided Markets

2008-01 * Fachgruppe Informatik: Jahresbericht 2007

2008-02 Henrik Bohnenkamp, Marielle Stoelinga: Quantitative Testing

2008-03 Carsten Fuhs, Jürgen Giesl, Aart Middeldorp, Peter Schneider-Kamp, René Thiemann, Harald Zankl: Maximal Termination

2008-04 Uwe Naumann, Jan Riehme: Sensitivity Analysis in Sisyphe with the AD-Enabled NAGWare Fortran Compiler

2008-05 Frank G. Radmacher: An Automata Theoretic Approach to the Theory of Rational Tree Relations

2008-06 Uwe Naumann, Laurent Hascoet, Chris Hill, Paul Hovland, Jan Riehme, Jean Utke: A Framework for Proving Correctness of Adjoint Message Passing Programs

2008-07 Alexander Nyßen, Horst Lichter: The MeDUSA Reference Manual, Second Edition

2008-08 George B. Mertzios, Stavros D. Nikolopoulos: The $\lambda$-cluster Problem on Parameterized Interval Graphs

2008-09 George B. Mertzios, Walter Unger: An optimal algorithm for the k-fixed-endpoint path cover on proper interval graphs

2008-10 George B. Mertzios, Walter Unger: Preemptive Scheduling of Equal-Length Jobs in Polynomial Time

2008-11 George B. Mertzios: Fast Convergence of Routing Games with Splittable Flows

2008-12 Joost-Pieter Katoen, Daniel Klink, Martin Leucker, Verena Wolf: Abstraction for stochastic systems by Erlang's method of stages

2008-13 Beatriz Alarcón, Fabian Emmes, Carsten Fuhs, Jürgen Giesl, Raúl Gutiérrez, Salvador Lucas, Peter Schneider-Kamp, René Thiemann: Improving Context-Sensitive Dependency Pairs

2008-14 Bastian Schlich: Model Checking of Software for Microcontrollers

2008-15 Joachim Kneis, Alexander Langer, Peter Rossmanith: A New Algorithm for Finding Trees with Many Leaves

2008-16 Hendrik vom Lehn, Elias Weingärtner and Klaus Wehrle: Comparing recent network simulators: A performance evaluation study

2008-17 Peter Schneider-Kamp: Static Termination Analysis for Prolog using Term Rewriting and SAT Solving

2008-18 Falk Salewski: Empirical Evaluations of Safety-Critical Embedded Systems

2008-19 Dirk Wilking: Empirical Studies for the Application of Agile Methods to Embedded Systems

2009-02 Taolue Chen, Tingting Han, Joost-Pieter Katoen, Alexandru Mereacre: Quantitative Model Checking of Continuous-Time Markov Chains Against Timed Automata Specifications

2009-03 Alexander Nyßen: Model-Based Construction of Embedded Real-Time Software - A Methodology for Small Devices

2009-04 Daniel Klünder: Entwurf eingebetteter Software mit abstrakten Zustandsmaschinen und Business Object Notation

2009-05 George B. Mertzios, Ignasi Sau, Shmuel Zaks: A New Intersection Model and Improved Algorithms for Tolerance Graphs

2009-06 George B. Mertzios, Ignasi Sau, Shmuel Zaks: The Recognition of Tolerance and Bounded Tolerance Graphs is NP-complete

2009-07 Joachim Kneis, Alexander Langer, Peter Rossmanith: Derandomizing Non-uniform Color-Coding I

2009-08 Joachim Kneis, Alexander Langer: Satellites and Mirrors for Solving Independent Set on Sparse Graphs

2009-09 Michael Nett: Implementation of an Automated Proof for an Algorithm Solving the Maximum Independent Set Problem

2009-10 Felix Reidl, Fernando Sánchez Villaamil: Automatic Verification of the Correctness of the Upper Bound of a Maximum Independent Set Algorithm

2009-11 Kyriaki Ioannidou, George B. Mertzios, Stavros D. Nikolopoulos: The Longest Path Problem is Polynomial on Interval Graphs

2009-12 Martin Neuhäußer, Lijun Zhang: Time-Bounded Reachability in Continuous-Time Markov Decision Processes

2009-13 Martin Zimmermann: Time-optimal Winning Strategies for Poset Games

2009-14 Ralf Huuck, Gerwin Klein, Bastian Schlich (eds.): Doctoral Symposium on Systems Software Verification (DS SSV'09)

2009-15 Joost-Pieter Katoen, Daniel Klink, Martin Neuhäußer: Compositional Abstraction for Stochastic Systems

2009-16 George B. Mertzios, Derek G. Corneil: Vertex Splitting and the Recognition of Trapezoid Graphs

2009-17 Carsten Kern: Learning Communicating and Nondeterministic Automata

2009-18 Paul Hänsch, Michaela Slaats, Wolfgang Thomas: Parametrized Regular Infinite Games and Higher-Order Pushdown Strategies

2010-02 Daniel Neider, Christof Löding: Learning Visibly One-Counter Automata in Polynomial Time

2010-03 Holger Krahn: MontiCore: Agile Entwicklung von domänenspezifischen Sprachen im Software-Engineering

2010-04 René Wörzberger: Management dynamischer Geschäftsprozesse auf Basis statischer Prozessmanagementsysteme

2010-05 Daniel Retkowitz: Softwareunterstützung für adaptive eHome-Systeme

2010-06 Taolue Chen, Tingting Han, Joost-Pieter Katoen, Alexandru Mereacre: Computing maximum reachability probabilities in Markovian timed automata

2010-07 George B. Mertzios: A New Intersection Model for Multitolerance Graphs, Hierarchy, and Efficient Algorithms

2010-08 Carsten Otto, Marc Brockschmidt, Christian von Essen, Jürgen Giesl: Automated Termination Analysis of Java Bytecode by Term Rewriting

2010-09 George B. Mertzios, Shmuel Zaks: The Structure of the Intersection of Tolerance and Cocomparability Graphs

2010-10 Peter Schneider-Kamp, Jürgen Giesl, Thomas Ströder, Alexander Serebrenik, René Thiemann: Automated Termination Analysis for Logic Programs with Cut

2010-11 Martin Zimmermann: Parametric LTL Games

2010-12 Thomas Ströder, Peter Schneider-Kamp, Jürgen Giesl: Dependency Triples for Improving Termination Analysis of Logic Programs with Cut

2010-13 Ashraf Armoush: Design Patterns for Safety-Critical Embedded Systems

2010-14 Michael Codish, Carsten Fuhs, Jürgen Giesl, Peter Schneider-Kamp: Lazy Abstraction for Size-Change Termination

2010-15 Marc Brockschmidt, Carsten Otto, Christian von Essen, Jürgen Giesl: Termination Graphs for Java Bytecode

2010-16 Christian Berger: Automating Acceptance Tests for Sensor- and Actuator-based Systems on the Example of Autonomous Vehicles

2010-17 Hans Grönniger: Systemmodell-basierte Definition objektbasierter Modellierungssprachen mit semantischen Variationspunkten

2010-18 Ibrahim Armaç: Personalisierte eHomes: Mobilität, Privatsphäre und Sicherheit

2010-19 Felix Reidl: Experimental Evaluation of an Independent Set Algorithm

2010-20 Wladimir Fridman, Christof Löding, Martin Zimmermann: Degrees of Lookahead in Context-free Infinite Games

2011-02 Marc Brockschmidt, Carsten Otto, Jürgen Giesl: Modular Termination Proofs of Recursive Java Bytecode Programs by Term Rewriting

2011-03 Lars Noschinski, Fabian Emmes, Jürgen Giesl: A Dependency Pair Framework for Innermost Complexity Analysis of Term Rewrite Systems

2011-04 Christina Jansen, Jonathan Heinen, Joost-Pieter Katoen, Thomas Noll: A Local Greibach Normal Form for Hyperedge Replacement Grammars

2011-11 Nils Jansen, Erika Ábrahám, Jens Katelaan, Ralf Wimmer, Joost-Pieter Katoen, Bernd Becker: Hierarchical Counterexamples for Discrete-Time Markov Chains

* These reports are only available as a printed version.

Please contact biblio@informatik.rwth-aachen.de to obtain copies.